\documentclass[11pt]{article}
\usepackage[margin=1in]{geometry}
\usepackage{amsmath,amsthm,amssymb}
\usepackage{graphicx}
\usepackage{wasysym}
\usepackage[linesnumbered,ruled,vlined]{algorithm2e}
\usepackage{hyperref}

{\bfseries}{\rmfamily}
\newtheorem{theorem}{Theorem}[section]
\newtheorem{definition}[theorem]{Definition}

\newtheorem{observation}[theorem]{Observation}
\newtheorem{corollary}{Corollary}[theorem]
\newtheorem{lemma}[theorem]{Lemma}
\newtheorem{proposition}[theorem]{Proposition}

\usepackage{xcolor}
\definecolor{light-gray}{gray}{0.95}

\usepackage{array}
\newcolumntype{L}{>{$}l<{$}}

\newcommand{\N}{\mathbb{N}} 
\newcommand{\Q}{\mathbb{Q}} 
\newcommand{\R}{\mathbb{R}} 
\newcommand{\Z}{\mathbb{Z}}
\newcommand{\E}{\mathbf{E}}
\newcommand{\bbP}{\mathbb{P}}
\newcommand{\bfP}{\mathbf{P}}
\newcommand{\calC}{\mathcal{C}}
\newcommand{\calS}{\mathcal{S}}
\newcommand{\calT}{\mathcal{T}}
\newcommand{\calA}{\mathcal{A}}
\newcommand{\calB}{\mathcal{B}}
\newcommand{\F}{\mathbb{F}}
\newcommand{\calQ}{\mathcal{Q}}
\newcommand{\PG}{\mathrm{PG}}
\newcommand{\com}{\mathrm{com}}
\newcommand{\spa}{\mathrm{span}}
\newcommand{\slsh}{\mathrm{slash}}
\newcommand{\load}{\mathrm{load}}
\newcommand{\msg}{\mathrm{msg}}

\usepackage{subfiles}

\begin{document}

\title{Optimal Multilevel Slashing for Blockchains}

\author{
Kenan Wood\thanks{Corresponding author}\\
\texttt{kewood@davidson.edu}\\
Davidson College
\and
Hammurabi Mendes\\
\texttt{hamendes@davidson.edu}\\
Davidson College
\and
Jonad Pulaj\\
\texttt{jopulaj@davidson.edu}\\
Davidson College
}

\date{}



\maketitle

\begin{abstract}
We present the notion of \emph{multilevel slashing}, where proof-of-stake blockchain validators can obtain gradual levels of assurance that a certain block is bound to be finalized in a global consensus procedure, unless an increasing and optimally large number of Byzantine processes have their staked assets \emph{slashed}---that is, deducted---due to provably incorrect behavior. Our construction is a highly parameterized generalization of combinatorial intersection systems based on finite projective spaces, with asymptotic high availability and \emph{optimal} slashing properties. Even under weak conditions, we show that our construction has asymptotically optimal slashing properties with respect to message complexity and validator load; this result also illustrates a fundamental trade off between message complexity, load, and slashing. In addition, we show that \emph{any} intersection system whose ground elements are disjoint \emph{subsets} of nodes (e.g. ``committees'' in committee-based consensus protocols) has asymptotic high availability under similarly weak conditions. Finally, our multilevel construction gives the flexibility to blockchain validators to decide how many ``levels'' of finalization assurance they wish to obtain. This functionality can be seen either as (i) a form of an early, slashing-based block finalization; or (ii) a service to support reorg tolerance.
\end{abstract}
\section{Introduction}

Blockchains are distributed systems with the task of (i) collecting concurrent \emph{transactions} that originate from its users; (ii) order these transactions into a history, which is often expressed as a linear sequence of \emph{blocks}, with each block containing an internal sequence of transactions; and (iii) maintain such history stored into a permanent, distributed ledger that reflects a global system state given the history. The ledger state could represent monetary balances or even the state of replicated programs (smart contracts) that execute in a virtual machine collectively simulated by participant nodes \cite{bitcoin, ethereum, avalanche, algorand, tezos, cardano}. Throughout this paper, we may refer to participant nodes, processes, and blockchain validators interchangeably.

Deciding how to group transactions into blocks and how to order blocks into the blockchain ledger is an application of Byzantine consensus, typically with some other properties and requirements in place, including: (i) participants always use digital signatures when they interact with the system; (ii) participation is dynamic, meaning that nodes join and leave the system at undetermined times; and (iii) in certain cases, participation is \emph{permissionless}, meaning that certain system actions (say, performing transactions or even participating in the consensus itself) require no previous global registration or identity-based approval. For example, Bitcoin participation is permissionless as any node that can produce a token demonstrating the completion of a certain computationally-expensive hashing task is allowed to participate in the consensus protocol, regardless of that node's identity. This is called a \emph{proof-of-work} blockchain. Note that any two nodes could produce the tokens mentioned above roughly at the same time, so the ledger can \emph{fork} during the system execution, and participants then heuristically define the longest chain of blocks as the authoritative one. In contrast, systems such as~\cite{ethereum, avalanche, algorand, tezos, cardano} work overall as follows. Discrete time \emph{slots} are defined, each associated with a \emph{committee} of participants. At each slot, the corresponding committee produces or acknowledges the next block of transactions by having some well-defined fraction of its participants to \emph{attest}---that is, vote---for such next block. If a Byzantine committee participant votes for two conflicting blocks (on different branches), which is prohibited by the voting mechanics, the participant is subject to \emph{slashing}: the participant's pre-deposited assets (called its \emph{stake}) are penalized, and the deducted penalty is distributed to other participants in the system. Having such pre-deposited assets is a prerequisite for participating in committees (and thus in the consensus protocol), and slashing is the incentive for nodes to behave correctly. This setting characterizes a \emph{proof-of-stake} blockchain.

An operational advantage of using committees rather than having a global vote procedure is that committees can initially restrict expensive communication primitives \cite{SriTouRB, Bracha, RSPBook} to its own participants, and later generate a compact \emph{committee signature}  that indicates internal agreement on a certain next block $v$ (say, using a threshold signature scheme such as~\cite{threshold1, threshold2}). Those compact committee signatures can then be communicated globally with the intention to reach global consensus on $v$.

Just as in proof-of-work systems, the ledger in proof-of-stake systems can also fork because committees can be temporarily or permanently isolated from the network due to technical outages, making attestations arrive asynchronously in different parts of the network, and thus creating multiple descendants of a given block. Hence, it is common to have a heuristic-based \emph{fork-choice rule} that constantly defines the ``best chain'' of ongoing operations\footnote{Just like Bitcoin does, except that Bitcoin's heuristic is simply defined as the longest chain.}, along with a separate \emph{finalization gadget}, which chooses one unique, canonical chain to be ``final'' \cite{ebb-flow, casper-ffg, grandpa-polkadot}. The blockchain literature often refers to the fork-choice's ``best chain'' as the \emph{available chain}, because its relatively simple heuristics allow applications to identify it quickly, thus settle quickly on the current state of the system. That is in contrast with what is often referred as the \emph{final chain}, which has been subject to the finalization gadget, and often depends on partially-synchronous assumptions for progress (\cite{pbft, hotstuff} among many others).

\textbf{The problem we solve} is that blockchain applications often need to know whether a transaction is ``confirmed'' (perhaps to settle a sale), which requires (a) observing a ledger block that includes such transaction; and (b) obtaining some guarantee that this block will be finalized later. However, currently, applications can either (i) \emph{quickly} identify the transaction in the available chain, but have no guarantees that this chain will eventually be finalized; or (ii) \emph{slowly} identify the transaction in the finalized chain, but be subject to an infeasible wait time that might completely break the application's functional requirements (user satisfaction, quick response to financial events, etc).

\textbf{Our solution} creates a ``sliding window'' in the latency-trust spectrum in settling transactions, and the applications can tradeoff speed and certainty according to their very particular functional requirements. In other words, applications will not only have the fork-choice rule (a temporary ``accounting mechanism'') or the finalization event (the permanent but slowly-moving global consensus) to deem that a particular block (or a particular transaction therein) is ``confirmed''. Specifically, we create a distributed mechanism where increasing \emph{levels} of trust can be obtained at increasing latency costs by querying other participant nodes as well as passively observing network events. Importantly, our design does not introduce any central control or ``hotspots'', as it is defined using the highly symmetrical mathematical structure of projective spaces over finite fields.

In essence, our construction can be interpreted as an intermediary, flexible \emph{trust} phase between the initial attestation tallying and the finalization. In this phase, a participant can obtain information from multiple sets of validators---which we call \emph{quorums}---indicating that a block $v$ is about to be finalized. If another participant obtains the same information for a different block $v'$, the intersection property of our structural construction will result in a significant amount of funds being slashed from the adversary. Importantly, we give the applications the flexibility to choose the balance of trust and potential adversarial slashing with our construction.

We note that this functionality, if further integrated into a blockchain system, can also be seen as (i) a form of an early, slashing-based \emph{block finalization}, as the block confirmation guarantees are now much more continuous between the quick, yet unreliable, fork-choice rule and the slow, yet reliable, finalization; or (ii) a service for \emph{reorg tolerance}. Reorgs \cite{schwarz2022three, neu2022two} are situations where applications consider some chain $v$ as the logical continuation of the blockchain ledger because a fraction of nodes attest $v$, but later are forced to consider $v'$ as such because a (typically larger) fraction of nodes became visible while attesting $v'$. In our system, once applications obtain a certain number of levels of assurance that $v$ is the next block to be finalized, a different block $v'$ that takes its place will incur significant adversarial slashing, optimal with respect to the magnitude of assurance levels originally obtained.

\textbf{Our technical contributions} are described below at a high level, but with pointers to the sections where we present our concrete constructions and proofs.
\begin{enumerate}
    \item We design a distributed architecture to support multilevel slashing by applying projective spaces over finite fields to committee-based consensus (Section \ref{Sec-SystemDesign}), a generalization of a previous approach that only used projective \emph{planes} \cite{rangarajan1995fault} in a context where slashing was irrelevant.
    \item We define and analyze \emph{slashability}---the relation of the query size/time and the magnitude of slashing associated with a level of trust. In particular, we show that our construction is \emph{optimal} with respect to worst-case message complexity and validator load, demonstrating a fundamental trade off between slashability, message complexity, and load (Section \ref{Sec-Slashing}).
    \item We prove that a \emph{general} class of similarly-designed intersection systems based on disjoint subsets of elements achieve asymptotic high availability under reasonable conditions (Section \ref{Sec-Availability}).
\end{enumerate}

Our construction creates an intersection pattern among sets of blockchain nodes in a manner that is reminiscent of quorum systems~\cite{naor, malkhi1998byzantine, qs-probabilistic, qs-permissionless, qs-general} (among others, discussed in Section~\ref{Sec-Related_Works}), but our purpose---and design---are not the same. Specifically, the mathematical intersections among sets of nodes intentionally uses a projective-space-based construction in order to define an \emph{additional} level of transaction confirmation on top of \emph{existing} blockchains that follow the availability-finality paradigm of~\cite{ebb-flow}. We note that obtaining the higher-dimensional structures used to define our quorums is expensive, but can be done \emph{a priori} for reasonable parameters, and later mapped to a running system, which we consider viable in practice. We discuss some practical scenarios in Section~\ref{Sec-SystemDesign}.

In addition to the core technical sections pointed to above, we include background on intersection systems and projective spaces in Section~\ref{Sec-Background}; we present our system design concretely in Section~\ref{Sec-SystemDesign}; we discuss related work in Section~\ref{Sec-Related_Works}; and we conclude with final remarks in Section~\ref{Sec-Conclusion}.

\section{Intersection Systems and Projective Spaces}
\label{Sec-Background}

In this section, we present basic definitions on intersection systems and projective spaces over finite fields, used in our intersection system construction, provided in Section~\ref{Sec-SystemDesign}. In this paper we use the following standard combinatorial notation: for a positive integer $m$, we denote by $[m]$ the set $\{1, \dots, m\}$. For probability computations, $\bfP(A)$ denotes the probability of an event $A$ and $\E X$ denotes an expectation of a random variable $X$.


\subsection{Intersection Systems}
Let us start with a definition of intersection system, below:
\begin{definition}[Intersection System]\label{definition-quorum-system}
    An \emph{intersection system} is simply a nonempty finite collection of finite sets $\mathcal{Q}$ such that any two sets $A,B \in \mathcal{Q}$ have a nonempty intersection. The sets in $\mathcal{Q}$ are called \emph{quorums}, and we refer to $\bigcup \mathcal{Q}$ as the \emph{ground set} of $\mathcal{Q}$.
\end{definition}

Intersection systems provide a framework for ensuring trust among decided or finalized blocks. 
Let $\mathbb{P}$ be a set of $n$ processes.
Suppose the ground set of an intersection system $\mathcal{Q}$ is the set of processes $\mathbb{P}$. Then if all processes in some quorum $A \in \mathcal{Q}$ attest to a block $v$ and all processes in some quorum $B \in \mathcal{Q}$ attest to $v' \ne v$, all processes will eventually learn this information (since every message is attached with a digital signature and the network is partially synchronous). We will therefore be able to deduce that every process in the nonempty set $A \cap B$ attested to different blocks and can slash these processes' staked assets. 
It is important to note that honest validators (those that do not attest to different blocks) never have their stake slashed with this protocol, even if they attest to a block that is not finalized.

Now, observe that if every quorum contains an adversarial process, these processes can simply be silent forever, which means that no decision can ever be made with this protocol, even if every correct process attested to the same block. This motivates Definition \ref{definition-failure-in-quorum-systems}. 

Note that Definitions \ref{definition-failure-in-quorum-systems} and \ref{definition-load} are similar to concepts in \cite{naor}, but their context is on replicated databases, not blockchain applications.
\begin{definition}[Availability]\label{definition-failure-in-quorum-systems}
    Let $\mathcal{Q}$ be an intersection system.
    Give each element of $\bigcup \Q$ a fixed probability of availability $p$, so the elements are independently non-faulty with probability $p$ and faulty (Byzantine) with probability $1-p$.
    Let $F_p(\mathcal{Q})$ denote the probability that every quorum in $\mathcal{Q}$ has at least one faulty element. The quantity $A_p(\mathcal{Q}) := 1-F_p(\mathcal{Q})$ is called the \emph{availability} of $\mathcal{Q}$ with respect to $p$.
\end{definition}

Another potential problem is that it is possible for two processes to trust different blocks if every process in $A \cap B$ is Byzantine, for quorums $A, B \in \mathcal{Q}$. Thus if $\min_{A, B \in \calQ} |A \cap B|$ is small, then it is possible for an adversary to make processes finalize different blocks with only a small amount of its stake being slashed.
Thus, a desirable property of intersection systems is that $|A \cap B|$ should be large for all $A, B \in \mathcal{Q}$.
\begin{definition}[Slashability]\label{definition-slashability}
    For an intersection system $\calQ$, define the \emph{slashability} of $\calQ$ to be the quantity $\min_{A, B \in \calQ} |A \cap B|$. The slashability of $\calQ$ is denoted $\slsh(\calQ)$.
\end{definition}

This definition of slashability is most relevant when validators have uniform stake. For highly heterogeneous validator stakes, committee-based constructions like ours in Section \ref{Sec-SystemDesign} can be adapted using techniques such as \emph{node virtualization}, where high-stake nodes ``simulate'' multiple nodes proportional to their stake.



In our design, the elements of the quorums are disjoint \emph{committees}, which are sets of processes in $\bbP$. 
\begin{definition}\label{definition-r-fraction}
    If $\mathbb{Q}$ is an intersection system whose elements are disjoint committees and $r \in (\frac{1}{2},1)$, we denote by $\bbP_r(\mathbb{Q})$ the intersection system
    \[
    \bbP_r(\mathbb{Q}) = \left\{S \subseteq \bbP: \exists \mathcal{Q} \in \mathbb{Q}, \forall Q \in \mathcal{Q}, |Q \cap S| \ge r|Q|\right\}.
    \]
    That is, a set of processes forms a quorum in $\bbP_r(\mathbb{Q})$ if it contains at least an $r$-fraction of every committee inside a quorum $\mathcal{Q} \in \mathbb{Q}$. The number $r$ is said to be the \emph{threshold} of $\bbP_r(\Q)$.
\end{definition}


The following definitions have fundamental connections to slashability, as seen in Section \ref{Sec-Slashing}.
In our system, we assume that processes \emph{actively} participate in obtaining quorums to reduce message complexity. In particular, committees select quorums uniformly at random to which they query messages. Thus, having small quorums is necessary, motivating the following definition.
\begin{definition}[Message Complexity]\label{definition-message-complexity}
    Given an intersection system $\Q$, let the maximum size of a quorum in $\Q$ be called the \emph{message complexity} of $\Q$, denoted $\msg(\Q)$.
\end{definition}

Additionally, with this system of actively obtaining quorums, we would like to ensure that no committee is overly busy handling queries, motivating another concept:
\begin{definition}[Load]\label{definition-load}
    Given an intersection system $\Q$, the \emph{load} of some $C \in \bigcup \Q$, denoted $\load_\Q(C)$, is the probability that a quorum of $\Q$ selected uniformly at random contains $C$. The \emph{load} of $\Q$ is defined to be the maximum load of any element of $\bigcup \Q$: $\load(\Q) := \max_{C \in \bigcup \Q} \load_\Q(C)$.
\end{definition}
When $\Q$ is clear from context, we simply write $\load(C)$ instead of $\load_\Q(C)$, where $C \in \bigcup \Q$.
There is a connection between the load of an element of an intersection system and the degree of that element (using terminology from graph theory).
\begin{definition}[Degree]\label{definition-degree}
    Given an intersection system $\Q$, the \emph{degree} of some $C \in \bigcup \Q$ in $\Q$, written $\deg_\Q(C)$, is the number of quorums of $\Q$ containing $C$.
    The maximum degree of any element of $\bigcup \Q$ is denoted $\Delta(\Q)$.
\end{definition}

When $\Q$ is clear from context, we write $\deg(C)$ instead of $\deg_\Q(C)$, where $C \in \bigcup \Q$. 
The following result should be clear from the definition of load and uniform selection.
\begin{observation}\label{observation-load-degree}
    If $\Q$ is an intersection system and $C \in \bigcup \Q$, then 
    \[
    \load(C) = \frac{\deg(C)}{|\Q|} \qquad \text{ and } \qquad \load(\Q) = \frac{\Delta(\Q)}{|\Q|}.
    \]
\end{observation}

\subsection{Projective Spaces}
Projective geometry provides a rich source of intersection systems that are highly symmetric (having a transitive automorphism group). Most of the definitions and notation in this section are similar to those presented in \cite{clark2012applications}.
To begin, it is known that finite fields have prime power order, and for each prime power $q$, there exists a unique finite field of order $q$, up to isomorphism.
Thus, given a prime power $q$, we may let $\F_q$ denote \emph{the} finite field of order $q$.
For the following definitions, let $V$ be a vector space over a field $\F$.
\begin{definition}[Projective Space]\label{definition-projective-space}
    The \emph{projective space} of $V$, denoted $\mathrm{PG}(V)$, is the set of 1-dimensional vector subspaces of $V$. In the case when $\F = \F_q$ for a prime power $q$ and $V = \F^{k+1}$, we may write $\PG(k, q)$ instead of $\PG(V)$. If $V$ is finite-dimensional, then the \emph{projective dimension} of $\PG(V)$ is $\dim \PG(V) = \dim V - 1$.
\end{definition}
\begin{definition}[Projective Subspace]\label{definition-projective-subspace}
    If $U$ is a vector subspace of $V$, then $\PG(U)$ is a \emph{projective subspace} of $\PG(V)$.
\end{definition}
\begin{definition}\label{definition-d-dim-projective-subspaces}
    If $d \ge 0$, let $\PG_d(V)$ be the set of all projective subspaces of $\PG(V)$ with (projective) dimension $d$. Just as before, if $\F = \F_q$ and $V = \F^{k+1}$, we write this as $\PG_d(k, q)$.
\end{definition}

The following result will be useful for analyzing the slashability of our system design.
\begin{proposition}\label{proposition-projective-subspace-intersections}
    Let $k \ge d \ge 0$ and let $q$ be a prime power. Then for any $S, T \in \PG_d(k, q)$, then $S \cap T$ is a projective subspace of $\PG(k, q)$ of dimension at least $2d-k$. If $2d \ge k$, this bound is sharp for some $S, T \in \PG_d(k, q)$.
\end{proposition}
\begin{proof}
    Please refer to Appendix~\ref{app-projective-geometry}.
\end{proof}
\begin{corollary}
\label{corollary-pg-quorum-system}
    If $2d > k$ and $q$ is any prime power, then $\PG_d(k, q)$ is an intersection system.
\end{corollary}

The following is a known result.
Given nonnegative integers $r,s$ and $q \ge 2$, recall the known \emph{$q$-Gaussian binomial coefficient} by the following equality, which yields Proposition~\ref{pg-size}.
\[
\binom{s}{r}_q = \frac{(q^s-1)(q^s-q) \cdots (q^s-q^{r-1})}{(q^r-1)(q^r-q) \cdots (q^r - q^{r-1})}.
\]
\begin{proposition}
\label{pg-size}
    For all $k \ge d \ge 0$ and prime powers $q$, we have
    \[
    |\PG_d(k, q)| = \binom{k+1}{d+1}_q \qquad \text{and} \qquad |\PG(k, q)| = \frac{q^{k+1}-1}{q-1}.
    \]
\end{proposition}


\section{System Design}

\label{Sec-SystemDesign}

Consider a system of $n$ processes $\bbP = \{P_1, \dots, P_n\}$. 
We assume that processes are non-faulty independently with probability $p$ (and faulty with probability $1-p$); similar failure models have been used in \cite{naor, rangarajan1995fault}. For practical applications, we use values $p$ of the form $\frac{2}{3} + \epsilon$ for a small $\epsilon$, so that the probability that at least $1/3$ of nodes display Byzantine behavior is negligible (so basic network primitives such as \cite{Bracha, SriTouRB} work).
We assume authenticated channels, that is, every message sent has a digital signature for which it is computationally infeasible for an adversary to forge.

We now describe a multilevel intersection system where, each level has an intersection system of its own with ground set composed of committees in $\bbP$. Obtaining a quorum asserting block $v$ within each level will increase the assurance that a $v$ is bound to be finalized in the global consensus---that is, increase the associated slashing in case $v$ is not finalized. We do this while allowing small quorums relative to system size (thus less communication complexity) and very high slashing relative to the size of the quorums. Our system also ensures a small load, as every committee is in the same number of quorums and no committee is particularly over-represented.
%
%
\textbf{Our construction is defined as follows.}
\begin{enumerate}
    \item
    Assume a set-up procedure to generate $|\PG(k, q)|$ \emph{committees} that equitably partition\footnote{An \emph{equitable} partition of a finite set $S$ is a partition of $S$ such that the sizes of the sets in the partition differ by at most 1.} the set of processes $\bbP$, where $k \ge 0$ is an integer and $q$ is a prime power such that $n \ge |\PG(k, q)|$. Note that when $q$ is a prime power, this is always well-defined. Denote the set of these committees by $\calC$, so that $|\calC| = |\PG(k, q)|$.

    \item
    Next, define a one-to-one correspondence $\com: \PG(k, q) \to \calC$, and two weakly increasing sequences with length $\ell$, the total number of levels:
\begin{itemize}
    \item $(d_j)_{j \in [\ell]}$ of integers in $\left(\frac{k}{2}, k\right)$;
    \item $(r_j)_{j \in [\ell]}$ of real numbers in $\left( \frac{1}{2}, p \right)$.
\end{itemize}

    \item
    For each level $j \in [\ell]$, the \emph{$j$th level committee intersection system} is defined to be
    \[
    \Q_j = \{\com(S): S \in \PG_{d_j}(k, q)\}.
    \]

    \item
    Finally, for all $j \in [\ell]$, the \emph{$j$th level process intersection system} is defined as
    \[
    \calQ_j = \bbP_{r_j}(\Q_j).
    \]
\end{enumerate}

\begin{figure}[htb]
	\centering
	\includegraphics[width=\textwidth]{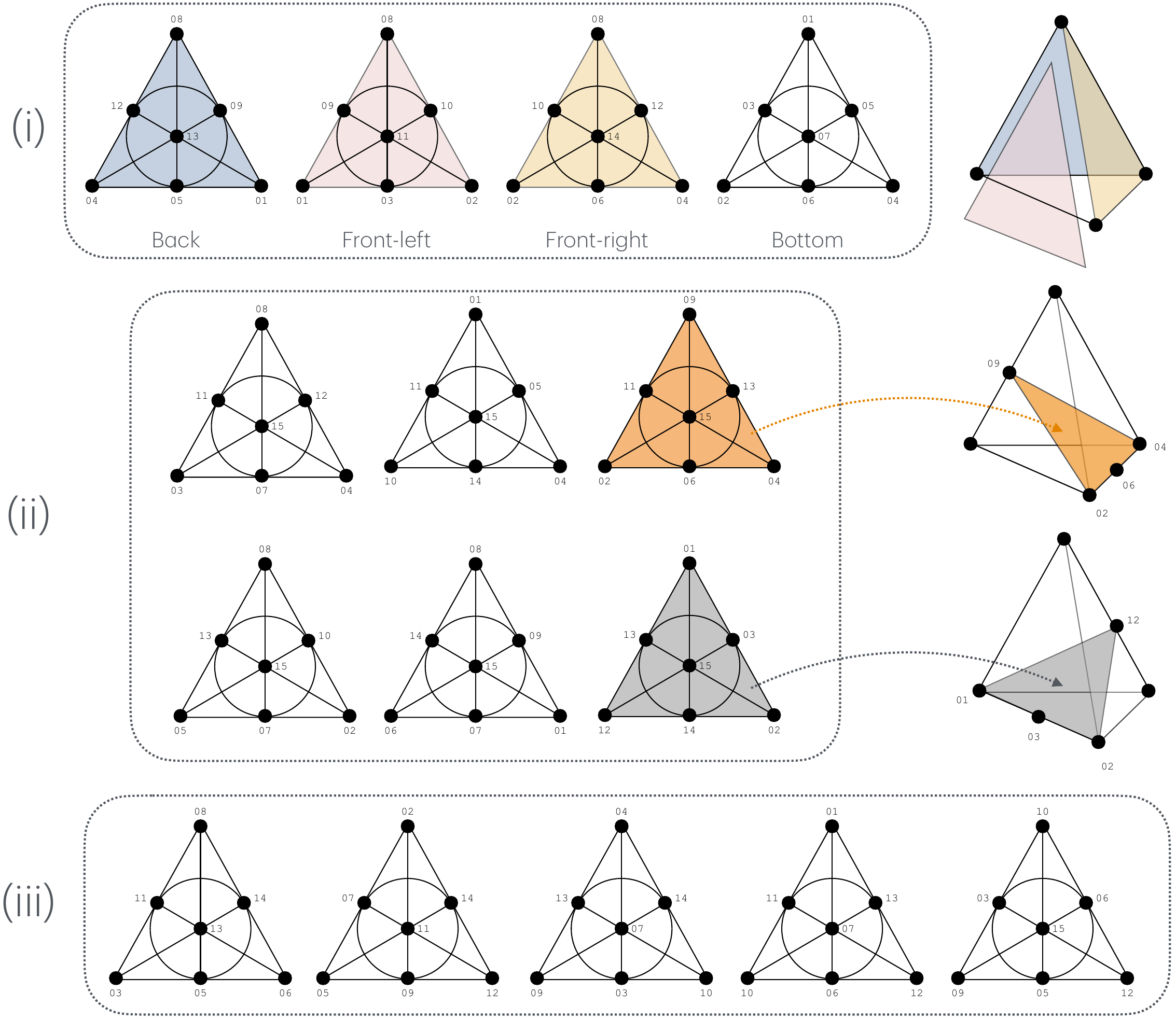}
	\caption{A tetrahedral visualization of $PG(3, 2)$, which contains 15 points, 15 planes, and 35 lines (circles are viewed as lines in projective geometry). More details in the text.}
	\label{fig-quorum-visualization}
\end{figure}
Notice that each $\Q_j$ is indeed an intersection system as $\PG_{d_j}(k, q)$ is an intersection system by Corollary \ref{corollary-pg-quorum-system}. In addition, observe that for all $1 \le i < j \le \ell$, every quorum in $\calQ_j$ contains a quorum in $\calQ_i$; since every $d_j$-dimensional projective subspace of $\PG(k, q)$ contains a $d_i$-dimensional projective subspace (this follows from $d_1 \le d_2 \cdots \le d_\ell$). Thus, we say that a process \emph{trusts a block $v$ with $j$ degrees of assurance} if that process obtains a quorum accepting $v$ from $\calQ_j$.

\textbf{Visualization.} To motivate the geometry of our construction with an example, in Figure~\ref{fig-quorum-visualization} we give a tetrahedral visualization of $PG(3,2)$, the smallest three-dimensional projective space. While the dimensions are low enough to allow a visualization, they only allow for a single level in our construction (with $d_1 =2$). Nevertheless, they should be useful to comprehend the system in higher dimensions. For the sake of simplicity, we disregard $r$ and focus on $\Q_1$. By definition $PG_2(3,2)$ is the set of planes in $PG(3,2)$, where each plane is isomorphic to the Fano plane. In (i), the outer faces of the tetrahedron are each a Fano plane. In (ii), the internal ``wedge'' planes are represented, with two planes highlighted. In (iii), we show Fano planes that are isomorphic to the additional planes inside the tetrahedron. Mapping the points in each Fano plane to corresponding points in the tetrahedron and preserving the incidence structure recovers the original plane. Our intersection system $\Q_1$ is set of all the visualized Fano planes. It is straightforward to check that any two distinct quorums intersect in a line, in other words they share three committees. 

\textbf{Choice of Parameters (Example).} It is crucial that parameters are chosen carefully, or the number of quorums at each level can quickly become too large. In that case, even the idea of precomputing quorums and later have them mapped to committees at runtime would be impractical. But many reasonable choices might exist: for example, consider a network of 2 million nodes, with committee sizes of about $8000$ nodes. We can set $k = 7, q = 2$ and have $d_1 = 4, d_2 = 5$, and $d_3 = 6$, which creates a 3-level intersection system with sizes  $|\Q_1| = \binom{8}{5}_2 = 97155$, $|\Q_2| = \binom{8}{6}_2 = 10795$ and $|\Q_3| = \binom{8}{7}_2 = 255$ (Proposition~\ref{pg-size}). In the first level, applications need to obtain assurances only from 4 committees forming a quorum ($d_1 = 4$) out of $97155$ quorums available ($|\Q_1|$). Any conflicting quorum intersects in $2^{2 \cdot 4 - 7 + 1} - 1 = 3$ committees in common  (Lemma~\ref{lemma-intersection-number}). If we assume that applications get threshold signatures from committees representing a fraction of $r = 60\%$ of their size, this must expose $3 \cdot (0.2 \cdot 8000)$ Byzantine nodes to slashing\footnote{Two subsets of $60\%$ of a ground set $S$ must intersect in $20\%$ of the nodes in $S$.}. In the second level, applications get \emph{one} extra committee ($d_2 = 5)$, essentially choosing one quorum out of $10795$ options  ($|\Q_2|$). Now, the number of committees in common jumps to $2^{2 \cdot 5 - 7 + 1} - 1 = 15$, thus exposing $15 \cdot (0.2 \cdot 8000)$ Byzantine nodes. In the third level, once again, applications get \emph{one} extra committee ($d_3 = 6$), essentially choosing one quorum out of $255$ options ($|\Q_3|$). Now, the number of committees in common jumps to $2^{2 \cdot 6 - 7 + 1} - 1 = 63$, thus exposing $63 \cdot (0.2 \cdot 8000)$ Byzantine nodes. Note how applications have many quorum choices at each level.


\textbf{Implementation.}
While the system as described above is mathematically elegant, and achieves asymptotically optimal slashing (Section~\ref{Sec-Slashing}) and high availability (Section \ref{Sec-Availability}), the above example points out that the number of quorums per level may become too large. As we (reasonably) assume that there is an operational network cost to keep track of quorums (for instance, joining gossip channels in \cite{gossipsub}), the applicability is compromised. In addition, calculating such large sets would be expensive. A probabilistic solution for reducing the number of quorums that works for \emph{reasonable} choices of parameters is the following (reasonable meaning that committee size is large enough so that $k, q$ are small enough).

The first two steps from the above construction remain the same, except, now we have additional parameters that heuristically bound the number of quorums in each level. Let $(\delta_j)_{j \in [\ell]}$ be a weakly increasing list of positive integers such that there exists $\delta_j$ distinct $d_j$-dimensional projective subspaces of $\PG(k, q)$ with a nonempty intersection. The construction of the $j$th level committee intersection system is now defined as follows. Define a random map $N_j: \PG(k, q) \to 2^{\PG_{d_j}(k, q)}$ by setting $N_j(A)$ to be a set of $\delta_j$ distinct $d_j$-dimensional projective subspaces of $\PG(k, q)$ that contain $A$, chosen uniformly at random\footnote{Given a set $S$, $2^S$ denotes the power set of $S$.}. Then, the $j$th level committee intersection system will be
\[
\Q_j' = \left\{\com(S): S \in \bigcup_{A \in \PG(k, q)} N_j(A) \right\},
\]
and the corresponding $j$th level process intersection system is still
\[
\calQ_j' = \bbP_{r_j}(\Q_j').
\]

With this construction, each committee is in \emph{approximately} the same number of quorums in $\Q_j'$ with high probability, and we always have that the size of each quorum in $\Q_j'$ equals the size of each quorum in $\Q_j$. Thus the load of $\Q_j'$ is approximately equal to the load of $\Q_j$, and the message complexities of $\Q_j'$ and $\Q_j$ are the same (refer to Definitions \ref{definition-load} and \ref{definition-message-complexity}, and Observation \ref{observation-load-degree}). We also know that $\Q_j' \subseteq \Q_j$ by construction, which immediately implies that the slashability of $\Q_j'$ (resp. $\calQ_j'$) is at least that of $\Q_j$ (resp. $\calQ_j$), and equal with very high probability. The availability results we prove in Section \ref{Sec-Availability} only depend on $\calC$, so we obtain the same results with $\calQ_j$ and with $\calQ_j'$. However, the number of quorums using this method is significantly reduced. In particular, by Proposition \ref{pg-size},
\begin{align*}
|\Q_j'| &= \left| \bigcup_{A \in \PG(k, q)} N_j(A) \right| \le \sum_{A \in \PG(k, q)} |N_j(A)| = \delta_j |\calC|\\
|\Q_j| &= \binom{k+1}{d_j+1}_q \approx q^{(k-d_j)(d_j+1)} \ge q^{kd_j - d_j} \approx |\calC|^{d_j - \frac{d_j}{k}} \ge |\calC|^{d_j-1}.
\end{align*}

Depending on the application, a more dense or less dense intersection system may be desirable. For example, the density (number of quorums out of all possible) may be related to resilience to an adversary that can target specific processes, instead of a case where Byzantine failures are independent and randomly distributed. However, analysis in different failure models is outside the scope of this paper and relegated to future work.
\section{Slashability}
\label{Sec-Slashing}
In this section, we show that our intersection system $\calQ_j$ has asymptotically optimal slashability, over a general class of intersection systems constructed from committees. We prove a formalization of the following: The slashability of the $j$th-level intersection system $\calQ_j = \bbP_{r_j}(\Q_j)$ is asymptotically optimal over all intersection systems built with the same set of committees and the same threshold $(r_j)$ with at least as good overall message complexity and load, allowing a certain trade off between the two quantities.


The proof of the statement above is given in Theorem \ref{theorem-slashing-optimality-formal}, following some useful lemmas. 
First, let us compute the size of the quorums in $\Q_j$ and the slashability of $\calQ_j$.

Consider the system with parameters as above, all viewed as a function of the number of processes $n$. Fix some $j \in [\ell]$ (also a function of $n$). For a more formalized treatment of asymptotic notations as in this section, see Section \ref{Sec-Availability}. Given functions $f$ and $g$ from $\N$ to $\R$, we say $f$ and $g$ are \emph{asymptotically equivalent}, written $f \sim g$, if $\lim_{n \to \infty} \frac{f(n)}{g(n)} = 1$.

These first two lemmas compute $\msg(\Q_j)$ and $\slsh(\Q_j)$.
\begin{lemma}\label{lemma-quorum-sizes}
    The size of each quorum in $\Q_j$ is precisely $\frac{q^{d_j+1}-1}{q-1}$. If $q = \omega(1)$, then this quantity is asymptotically equivalent to $q^{d_j}$.
\end{lemma}
\begin{proof}
    Consider any $Q \in \Q_j$. Then there exists some $S \in \PG_{d_j}(k, q)$ such that $\com(S) = Q$. This implies that $S$ is a $d_j$-dimensional projective subspace of $\PG(k, q)$. Hence $S$ is isomorphic to $\PG(d_j, q)$, which has exactly $\frac{q^{d_j+1}-1}{q-1}$ elements, by Proposition \ref{pg-size}. Since $\com$ is a bijection, it follows that $|S| = |Q| = \frac{q^{d_j+1}-1}{q-1}$. The second statement follows immediately.
\end{proof}

\begin{lemma}
    \label{lemma-intersection-number}
    The slashability of $\Q_j$ is precisely $\frac{q^{2d_j-k+1}-1}{q-1}$. If $q = \omega(1)$, then this quantity is asymptotically equivalent to $q^{2d_j-k}$.
\end{lemma}
\begin{proof}
    Suppose $Q, R \in \Q_j$. Then there exist $S, T \in \PG_{d_j}(k, q)$ where $\com(S) = Q$ and $\com(T) = R$. By Proposition \ref{proposition-projective-subspace-intersections}, $S \cap T$ is a projective subspace of $\PG(k, q)$ of dimension at least $2d_j-k$. By Proposition \ref{pg-size}, this implies that $|S \cap T| \ge \frac{q^{2d_j-k+1}-1}{q-1}$. Since $\com$ is a bijection, this implies that $\slsh(\Q_j) \ge \frac{q^{2d_j-k+1}-1}{q-1}$. However, since the bound in Proposition \ref{proposition-projective-subspace-intersections} is sharp, this shows $\slsh(\Q_j) = \frac{q^{2d_j-k+1}-1}{q-1}$. The second statement follows immediately.
\end{proof}



Then Lemmas \ref{lemma-quorum-sizes} and \ref{lemma-intersection-number} show that if $q = \omega(1)$,
\[
\slsh(\Q_j) \sim q^{2d_j-k} = (q^{d_j})^{2-\frac{k}{d_j}} \sim \msg(\Q_j)^{2-\frac{k}{d_j}}.
\]
Hence $\slsh(\Q_j) \sim \msg(\Q_j)^{2-\frac{k}{d_j}}$.

In the following result, we assume each committee has the same size, but loosening this restriction does not change the result and is only for simplification.
\begin{proposition}\label{proposition-slashability-calQ}
    Suppose every committee in $\calC$ contains exactly $c$ processes such that $r_j c \in \Z$. Then
    \[
    \slsh(\calQ_j) = (2r_j-1) c \cdot \slsh(\Q_j) = (2r_j-1) c \frac{q^{2d_j-k+1}-1}{q-1} = (2r_j-1)\frac{q^{2d_j-k+1}-1}{q^{k+1}-1} n.
    \]
    If $q = \omega(1)$,
    \[
    \slsh(\calQ_j) \sim (2r_j-1) q^{2d_j-2k} n
    \]
\end{proposition}
\begin{proof}
    Suppose $S, T \in \calQ_j$. Then there exists $\calS, \calT \in \Q_j$ such that $S$ contains at least $r_j c$ processes from each committee in $\calS$, and similarly for $T$. Then for each $C \in \calS \cap \calT$, we have
    \begin{align*}
    |S \cap T \cap C| &= |(S \cap C) \cap (T \cap C)|\\
    &= |S \cap C| + |T \cap C| - |(S \cap C) \cup (T \cap C)|\\
    &\ge r_jc + r_jc - c = (2r_j-1)c.
    \end{align*}
    Also, $|\calS \cap \calT| \ge \slsh(\Q_j)$. Hence
    \begin{align*}
        |S \cap T| &= \sum_{C \in \calC} |S \cap T \cap C|\\
        &\ge \sum_{C \in \calS \cap \calT} |S \cap T \cap C|\\
        &\ge (2r_j-1)c\cdot\slsh(\Q_j).
    \end{align*}
    It follows that $\slsh(\calQ_j) \ge (2r_j-1)c\cdot \slsh(\Q_j)$. 
    
    To prove equality, consider $\calS, \calT \in \Q_j$ such that $|\calS \cap \calT| = \slsh(\Q_j)$. In the following, we say that the \emph{identity} of process $P_i \in \bbP$ is $i$; the set of process identities is thus $[n]$. Let $S \subseteq \bbP$ be defined by the following: for all $X \in \mathcal{S}$, pick each of the $r_j c$ processes in $X$ with the least identities to be in $S$. Define $T \subseteq \bbP$ similarly, by selecting the $r_jc$ processes of each $X \in \calT$ with the greatest identities to be in $T$.
    Since $r_j > \frac{1}{2}$, for all $X \in \calS \cap \calT$, we have $(S \cap X) \cup (T \cap X) = X$, so that
    \begin{align*}
        |S \cap T \cap X| &= |S \cap C| + |T \cap C| - |(S \cap C) \cup (T \cap C)| = r_jc + r_jc - c = (2r_j-1)c.
    \end{align*}
    Hence
    \[
    |S \cap T| = \sum_{X \in \calS \cap \calT} |S \cap T \cap X| = (2r_j-1)c\cdot \slsh(\Q_j).
    \]
    It follows that $\slsh(\calQ_j) = (2r_j-1)c\cdot \slsh(\Q_j)$. 

    Since $\calC$ is a partition of $\bbP$, we know $c |\calC| = n$, so 
    \[
    c = \frac{n}{|\calC|} = \frac{n}{|\PG(k, q)|} = \frac{n(q-1)}{q^{k+1}-1},
    \]
    by Proposition \ref{pg-size}. Substituting this expression for $c$ applying Lemma \ref{lemma-intersection-number}, we obtain 
    \[
    \slsh(\calQ_j) = (2r_j-1) c \frac{q^{2d_j-k+1}-1}{q-1} = (2r_j-1)\frac{q^{2d_j-k+1}-1}{q^{k+1}-1} n.
    \]
    The simplified asymptotic expression for $\slsh(\calQ_j)$ when $q = \omega(1)$ follows immediately.
\end{proof}

Let us now show the asymptotic optimality of our system. Denote $m = |\calC|$, so $m$ is the number of committees. 
Recall Definitions \ref{definition-message-complexity} and \ref{definition-load}. The following defines the set of intersection systems over which we prove optimality.
\begin{definition}\label{definition-quorum-systems-optimality}
    Given $r \in (\frac{1}{2}, 1)$, $\lambda \in (0, 1)$, and a positive integer $1 \le \mu \le m$, let $\mathbb{S}(\mu, \lambda, r)$ be the set of all intersection systems of the form $\bbP_r(\Q)$, where $\Q$ is an intersection system with ground set $\calC$ such that the product of its message complexity and its load is at most $\mu \cdot \lambda$; that is, $\msg(\Q) \cdot \load(\Q) \le \mu \cdot \lambda$.
\end{definition}

Since we will prove the optimality of $\calQ_j$, we are interested in the maximum slashability of any intersection system in $\mathbb{S}(\msg(\calQ_j), \load(\Q_j), r_j)$. A special case of an intersection system in this set is $\bbP_{r_j}(\Q)$, where $\bigcup\Q = \calC$ and $\msg(\Q) \le \msg(\Q_j)$ and $\load(\Q) \le \load(\Q_j)$, although we prove optimality in a more general setting.

\begin{lemma}\label{lemma-slashing-upper-bound}
    Suppose each committee has size $c$ and $rc$ is an integer. Let $1 \le \mu \le m$ and $\lambda \in (0, 1)$.
    For all $\calQ \in \mathbb{S}(\mu, \lambda, r)$, we have $\slsh(\calQ) \le (2r-1) c \cdot \mu \cdot \lambda$.
\end{lemma}
\begin{proof}
    Let $\calQ \in \mathbb{S}(\mu, \lambda, r)$; let $\Q$ be an intersection system such that $\calQ = \bbP_r(\Q)$. Then $\msg(\Q) \cdot \load(\Q) \le \mu \cdot \lambda$. Select $\calA, \calB \in \Q$ independently and uniformly at random. Then $\slsh(\Q) \le |\calA \cap \calB|$, so $\slsh(\Q) \le \E|\calA \cap \calB|$. For each committee $C \in\bigcup \Q$, let $X_C$ be the indicator random variable for $C \in \calA$ and $C \in \calB$, so that $|\calA \cap \calB| = \sum_{C \in \calC} X_C$. By linearity of expectation,
    $
    \E|\calA \cap \calB| = \sum_{C \in \calC} \E X_C.
    $
    A simple double counting argument shows that
    $
    \sum_{C \in \calC} \deg(C) = \sum_{\calS \in \Q} |\calS| \le \msg(\Q) \cdot |\Q|
    $,
    therefore, by Observation \ref{observation-load-degree}, we have 
    $
    \sum_{C \in \calC} \load(C) \le \msg(\Q).
    $
    Also, $\load(C) \le \load(\Q)$ for all $C \in \calC$. Since for all $C \in \calC$, the events $C \in \calA$ and $C \in \calB$ are independent with probability $\load(C)$, it follows that 
    $
    \E X_C = \bfP(C \in \calA, C \in \calB) = \bfP(C \in \calA) \cdot \bfP(C \in \calB) = \load(C)^2.
    $
    Hence,
    \begin{align*}
        \slsh(\Q) &\le \E|\calA \cap \calB| = \sum_{C \in \calC} \load(C)^2
        \le \sum_{C \in \calC} \load(\Q) \cdot \load(C)
        \le \load(\Q) \cdot \msg(\Q) \le \mu \cdot \lambda.
    \end{align*}
    By the proof of Proposition \ref{proposition-slashability-calQ},
    $
    \slsh(\calQ) = (2r-1)c \cdot \slsh(\Q) \le (2r-1)c \cdot \lambda \cdot \mu.
    $
\end{proof}

We now come to our main result with respect to slashability, which is a formalization of the informal statements outlined in the beginning of this section.
\begin{theorem}\label{theorem-slashing-optimality-formal}
    Suppose each committee in $\calC$ contains $c$ processes and $r_j c \in \Z$. Then $\calQ_j \in \mathbb{S}(\msg(\Q_j), \load(\Q_j), r_j)$ and if $q = \omega(1)$, then
    \[
    \slsh(\calQ_j) \sim \max \left\{\slsh(\calQ): \calQ \in \mathbb{S}(\msg(\Q_j), \load(\Q_j), r_j)\right\}.
    \]
\end{theorem}
\begin{proof}
    It is clear by definition that $\calQ_j \in \mathbb{S}(\msg(\Q_j), \load(\Q_j), r_j)$. Suppose $q = \omega(1)$.
    It is also easy to see that
    \[
    \lim_{n \to \infty} \frac{\slsh(\calQ_j)}{\max \left\{\slsh(\calQ): \calQ \in \mathbb{S}(\msg(\Q_j), \load(\Q_j), r_j)\right\}} \le 1.
    \]
    By Proposition \ref{pg-size}, $\msg(\Q_j) = \frac{q^{d_j+1}-1}{q-1}$. Since every committee in $\calC$ has the degree $\Delta(\Q_j)$ in $\Q_j$ and every quorum has size $\msg(\Q_j)$, we also have
    \[
    m \cdot \Delta(\Q_j) = \sum_{C \in \calC} \deg_{\Q_j}(C) = \sum_{Q \in \Q_j} |Q| = |\Q_j| \cdot \msg(\Q_j).
    \]
    It follows by Observation \ref{observation-load-degree} that
    $
    \load(\Q_j) = \frac{\Delta(\Q_j)}{|\Q_j|} = \frac{\msg(\Q_j)}{m}
    $.
    
    Also, by Proposition \ref{pg-size}, $m = |\calC| = |\PG(k, q)| = \frac{q^{k+1}-1}{q-1}$. By Proposition \ref{proposition-slashability-calQ}, $\slsh(\calQ_j) = (2r_j-1) c \frac{q^{2d_j-k+1}-1}{q-1}$. By Lemma \ref{lemma-slashing-upper-bound}, we then have
    \begin{align*}
        \frac{\slsh(\calQ_j)}{\max \left\{\slsh(\calQ): \calQ \in \mathbb{S}(\msg(\Q_j), \load(\Q_j), r_j)\right\}} &\ge \frac{(2r_j-1)c\frac{q^{2d_j-k+1}-1}{q-1}}{(2r_j-1)c \cdot \msg(\Q_j) \cdot \load(\Q_j)}\\
        &= \frac{\frac{q^{2d_j-k+1}-1}{q-1}}{\frac{\msg(\Q_j)^2}{m}}\\
        &= \frac{q^{2d_j-k+1}-1}{q-1} \cdot \frac{\frac{q^{k+1}-1}{q-1}}{\left( \frac{q^{d_j+1}-1}{q-1} \right)^2}\\
        &= \frac{(q^{2d_j-k+1}-1)(q^{k+1}-1)}{(q^{d_j+1}-1)^2}
        \sim 1,
    \end{align*}
    as $n \to \infty$, where the last $1$ denotes the constant function on $\N$ that is always equal to the number $1$. It follows that
    $
    \slsh(\calQ_j) \sim \max \left\{\slsh(\calQ): \calQ \in \mathbb{S}(\msg(\Q_j), \load(\Q_j), r_j)\right\}.
    $
\end{proof}

\section{Availability}
\label{Sec-Availability}

In this section we demonstrate that under only mild assumptions, \emph{any} multilevel intersection system that generalizes our approach has an asymptotically high availability (including the case when the number of levels is 1). We consider the results of this section relevant on their own, thus the results will be presented with a more general notation (from Section~\ref{Sec-Background}), rather than relying on specific notation from of our concrete construction in Section \ref{Sec-SystemDesign}. With inspiration from some proof ideas seen in~\cite{rangarajan1995fault}, we prove a formalized version of the following:
    Suppose processes are available with probability $p$. Suppose $\calC$ is a partition of the processes into committees and $\Q$ is any intersection system with ground set $\calC$. If $\frac{1}{2} < r < p$ and the smallest committee has size $\Omega(\log n)$, then the intersection system $\bbP_r(\Q)$
    has availability converging to 1.

The formal statement and proof of this statement is concluded in Corollary~\ref{corollary-availability}, following some useful lemmas below. Lets start by formalizing our assumptions. For each $n \ge 1$, let $\calC^n$ be a partition of an $n$-element set of processes, and let $\Q^n$ be an intersection system with ground set $\calC^n$. Let $p$ be the probability a process is available in its steady-state, with $p > 1/2$. For each $n \ge 1$, let $r_n \in (\frac{1}{2}, p)$ so that $(r_n)_{n \in \N}$ is weakly increasing and $r := \sup_{n \in \N} r_n < p$.\footnote{Note that the $r$'s presented in this section are not related to the $\ell$ different $r$-values used to define our multilevel intersection system in Section \ref{Sec-SystemDesign}.} Finally, define $c_n = \min_{C \in \calC^n} |C|$; that is, $c_n$ the smallest size of any set within the partition $\calC^n$ (that, more generally, models committees).


Now let us bound the probability that a committee of size $c$ does not have at least $rc$ available processes. Denote this probability by $F_p(c; r)$. Define the function $a_1(x) = \frac{(p-x)^2}{2 - p - x}$ for $x \in [0, r]$.
\begin{lemma}\label{committee_failure}
    Suppose $0 < r < p < 1$. We have $F_p(c; r) \le e^{-a_1(r) \cdot c}$.
\end{lemma}
\begin{proof}
    Let $Z$ be the number of unavailable processes in a committee of size $c$. It is easy to see that $Z$ is a binomial random variable with parameters $c$ and $q = 1-p$. Then we have
    \[
    F_p(c; r) = \bfP(Z > c - rc) \le \bfP\left(Z \ge \frac{1-r}{q}\cdot cq\right) = \bfP\left( Z \ge \left(1+\frac{1-r-q}{q}\right) \cdot cq \right).
    \]
    Using $\delta = \frac{1-r-q}{q} = \frac{p-r}{q}$ in Lemma \ref{chernoff} from Appendix \ref{app-chernoff} ($\delta > 0$ since $r < p$), we obtain
    \[
    F_p(c; r) \le \exp\left(-\frac{1}{2+\frac{p-r}{q}}\left(\frac{p-r}{q}\right)^2 c q \right) = \exp\left( -\frac{(p-r)^2}{2 - p - r}\cdot c \right)
    \]
\end{proof}

\begin{lemma}\label{availability_bound}
    For all positive integers $n$, we have
    \begin{align*}
        A_p(\bbP_{r_n}(\Q^n)) &\ge 1 - \frac{n}{c_n} \cdot e^{-a_1(r)\cdot c_n}.
    \end{align*}
\end{lemma}
\begin{proof}
    Recall $r = \sup_{n \in \N} r_n < p$. Fix some positive integer $n$.
    For each $C \in \calC^n$, let $E_C$ be the event that at least $r_n|C|$ of the processes in $C$ are available. Then each $E_C$ is independent since the sets in $\calC^n$ are pairwise disjoint. Hence, by Lemma \ref{committee_failure},
    \begin{align*}
        A_p(\bbP_{r_n}(\Q^n)) &\ge \bfP\left(\bigcap_{C \in \calC^n} E_C\right) = \prod_{C \in \calC^n} \bfP(E_C)\\
        &= \prod_{C \in \calC^n} (1-F_p(|C|; r_n))
        \ge \prod_{C \in \calC^n} \left(1 - \exp\left( -a_1(r_n) \cdot |C| \right)\right).
    \end{align*}
    A straightforward computation shows that for $x \in [0,r]$, we have $a_1'(x) = \frac{(x-p)(4-3p-x)}{(2-p-x)^2}$; since $0 \le x \le r < p$, we have $x-p < 0$, $4-3p-x > 0$, and $(2-p-x)^2 > 0$, which implies $a_1'(x) < 0$ on $[0,r]$.
    Thus $a_1(x)$ is decreasing on $[0,r]$. Since $r_n \le r < p$, this shows $a_1(r_n) \ge a_1(r)$. Also, $|C| \ge c_n$ for all $C \in \calC^n$ implies $|\calC^n| \le \frac{n}{c_n}$. Putting these results together, we obtain
    \begin{align*}
        A_p(\bbP_{r_n}(\Q^n)) &\ge \prod_{C \in \calC^n} \left(1 - \exp(-a_1(r) \cdot c_n)\right)
        = \left( 1 - e^{-a_1(r) \cdot c_n} \right)^{|\calC^n|}\ge \left( 1 - e^{-a_1(r) \cdot c_n} \right)^{\frac{n}{c_n}}.
    \end{align*}
    Recall that for all $k \ge 1$
    and $0 \le y \le x \le 1$, the following known inequality $k(x-y) \ge x^k - y^k$ holds. Using $k = \frac{n}{c_n}$ (which is at least $1$ since $c_n \le n$), $x = 1$, $y = 1 - e^{-a_1(r) \cdot c_n}$ yields $\frac{n}{c_n} ( 1 - (1 - e^{-a_1(r) \cdot c_n}) ) \ge 1 - \left( 1 - e^{-a_1(r) \cdot c_n} \right)^{\frac{n}{c_n}}$, which, following algebraic manipulations, implies the desired bound.
\end{proof}

Now we are ready to prove our main theorem. We say a property $R(n)$ that is true or false for every positive integer $n$ holds \emph{for sufficiently large $n$} if there exists a positive integer $N$ such that $n \ge N$ implies $R(n)$ is true.
\begin{theorem}\label{high_availability}
    Suppose $a > 0$ such that for all sufficiently large $n$, we have $c_n \ge a \log n$. If $a = \frac{b+1}{a_1(r)} = \frac{2-p-r}{(p-r)^2}(b+1)$ for some constant $b$, then 
    \[
    A_p(\bbP_{r_n}(\Q^n)) \ge 1 - \frac{1}{a \cdot n^b \cdot \log n} = 1 - O\left( \frac{1}{n^b \cdot \log n}\right).
    \]
\end{theorem}
\begin{proof}
    Using the bound in Lemma \ref{availability_bound}, for sufficiently large $n$, we obtain
    \begin{align*}
        A_p(\bbP_{r_n}(\Q^n)) &\ge 1 - \frac{n}{c_n} \cdot e^{-a_1(r)\cdot c_n}
        \ge 1 - \frac{n}{a\log n} \cdot e^{-a_1(r) \cdot a \log n}\\
        &= 1 - \frac{n}{a \log n} \cdot n^{-a_1(r) a}
        = 1 - \frac{1}{a\cdot n^{a_1(r)a-1}\cdot \log n}
        = 1 - \frac{1}{a \cdot n^b \cdot \log n}.
    \end{align*}
\end{proof}

This immediately leads to the following corollary.
\begin{corollary}
\label{corollary-availability}
    Suppose $a$ and $b$ are constants satisfying the hypotheses of Theorem \ref{high_availability}. If $b \ge 0$, then
    $
    \lim_{n \to \infty} A_p(\bbP_{r_n}(\Q^n)) = 1.
    $
\end{corollary}
\section{Related Work}
\label{Sec-Related_Works}

It is common practice~\cite{ethereum, avalanche, algorand, tezos, cardano} for proof-of-stake blockchains to employ fork-choice rules in order to define an available chain, and rely in an additional ``finalization gadget'' to execute global consensus. In particular, we note the role of Casper in Ethereum~\cite{casper-ffg}, and point to a well-documented conceptual separation of available vs. final chains in~\cite{ebb-flow}. The later paper also points to the fact that this separation, besides practical, is also more general in the sense that a diverse set of consensus algorithms~\cite{pbft, hotstuff, streamlet} (among others) could be applicable. Reorg attacks~\cite{schwarz2022three, neu2022two} are also related to our work as our multilevel construction can precisely quantify a lower bound on slashing if a reorg were to take place---so our work could be seen as a tool to mitigate the functional effects of reorgs.

As noted in the introduction, our work is reminiscent to (and certainly it is inspired by) previous work in \emph{quorum systems}. Quorum systems are set structures whose elements are typically distributed processes (say, servers in a distributed system), so that any two such sets (called \emph{quorums}) intersect in at least one process. The idea is that applications can obtain an acknowledgment of an operation from all members of a chosen quorum, so that any two such acknowledgements are consistent, due to the intersecting property. Quorum systems have been studied extensively \cite{naor, qs-probabilistic, federated-byzantine}, with applications to consensus \cite{stellar}, database synchronization \cite{qs-databases}, finite-state-machine replication \cite{qs-fsm-datatypes}, mutual exclusion \cite{maekawa1985n}, among many others.

We are interested in some crucial metrics, such as server load and system availability, that are also found in the extensive literature of quorum systems, for example in \cite{naor, peleg1995availability}. We note that quorum system design is also quite diverse, being highly impacted by its target applications. For abstract-data-type replication, ADT-specific information can play a role \cite{qs-fsm-datatypes, qs-dynamic-size}. Another interesting application showing high impact on quorum design is federated Byzantine consensus: in this case, non-uniform quorum systems, where each participant has its own notion of membership, are studied in \cite{federated-byzantine}, with ideas originating from practical systems such as the Stellar consensus protocol \cite{stellar}. An additional important considerations for quorum design are whether the system allows dynamic participation \cite{qs-dynamic, qs-permissionless}. 

Block designs are well-studied structures in combinatorics \cite{colbourn2010crc} which provide a rich family of parameterized quorum systems \cite{colbourn2001quorum}. A well-known family of block designs corresponds to finite projective planes and their respective quorum systems. One of its earliest applications in distributed systems is an algorithm that uses $O(\sqrt{n})$ messages to create mutual exclusion in a network, where $n$ is the number or nodes \cite{maekawa1985n}. Although finite projective planes that yield small quorums are desirable due to reduction in communication complexity, asymptotically their availability goes to zero \cite{rangarajan1995fault, kumar1991high}. However, a construction with finite projective planes where points are disjoint subsets that cover the ground set and quorums are achieved on an $r$ fraction of elements in the subsets yields asymptotic high availability \cite{rangarajan1995fault}. A novel way of leveraging projective spaces in quorum systems for the in-network data storage paradigm is introduced and explored in \cite{sarkar2006double, luo2011geoquorum}, where a 2D network is ``lifted'' onto a sphere using a projective map to enable more flexible quorums.

\section{Conclusion}
\label{Sec-Conclusion}

In this work, we propose an application of combinatorial designs using projective spaces over finite fields to provide gradual levels of consensus---that is, getting gradual assurance that a certain block is bound to be finalized in a global consensus procedure. We combine our design with a committee-based approach, in order to provide high availability. In fact, we prove not only that our construction is subject to high availability, but we show that \emph{any} approach that uses our ``sliding window'' of obtaining attestations in the range strictly between $\frac{1}{2}$ and the individual process availability also forms a highly available intersection system. In addition, we demonstrate that our construction has optimal slashability---the extent we penalize an adversary's assets upon protocol non-compliance---compared to other systems that operate under similar load and message complexity.

We consider the potential applicability of our system very exciting, as  we envision that highly-connected participant nodes can even buy and sell ``trust certificates'' associated with assurances that a block is bound to be finalized. That is, nodes obtain collections of individual and threshold signatures that form multiple quorums, and then offer these trust certificates as \emph{insurance} to blockchain applications.

Having a more gradual consensus---a ``sliding window'' of trust---can enable large improvements in blockchain usability, providing services such as early slashing-based finalization, reorg tolerance, and even supporting transaction insurance for reorgs. We are excited to spearhead initial theoretical work in this direction.


\bibliography{Bibliography.bib}
\bibliographystyle{plainurl}

\newpage
\appendix
\section{Projective Geometry Proofs}
\label{app-projective-geometry}

\begin{proof}[Proof of Proposition \ref{proposition-projective-subspace-intersections}]
    Suppose $S, T \in \PG_d(k,q)$. Then there exists $(d+1)$-dimensional vector subspaces $U, W \subseteq \F_q^{k+1}$ such that $S = \PG(U)$ and $T = \PG(W)$. It follows that $S\cap T = \PG(U \cap W)$, so $S \cap T$ is indeed a projective subspace of $\PG(k, q)$. Since $U + W$ is a subspace of $\F_q^{k+1}$, $U+W$ has dimension at most $k+1$, which shows
    \begin{align*}
    \dim (S \cap T) &= \dim \PG(U \cap W) = \dim (U \cap W)-1\\
    &= \dim U + \dim W - \dim (U + W) - 1\\
    &\ge \dim U + \dim W - (k+1) - 1\\
    &= 2(d+1) - k - 2 = 2d - k.
    \end{align*}

    To show that this bound is sharp for some choice of $S$ and $T$, suppose $2d \ge k$ and consider any basis $v_1, \dots, v_{k+1}$ of $\F_q^{k+1}$. Let 
    \[
    U = \spa(v_1, \dots, v_{d+1}) \quad \text{and} \quad W = \spa(v_{k+1}, v_{k}, \dots, v_{k-d+1}).
    \]
    Clearly both $U$ and $W$ are vector spaces of dimension $d+1$. Also, 
    \[
    U \cap W = \spa(v_{k-d+1}, \dots, v_{d+1}),
    \]
    which is a vector space of dimension $(d+1)-(k-d+1)+1 = 2d-k+1$. Hence, $\PG(U), \PG(W) \in \PG_d(k, q)$ and $\PG(U) \cap \PG(W) = \PG(U \cap W)$ is a projective subspace of $\PG(k, q)$ with projective dimension $\dim(U \cap W) - 1 = 2d-k$, as desired.
\end{proof}
\section{Chernoff Bound}
\label{app-chernoff}
\begin{lemma}[Chernoff Bound]\label{chernoff}
    Let $Z$ be a binomial random variable with parameters $n$ and $q$. If $\delta > 0$, then
    \[
    \bfP(Z \ge (1+\delta) n q) \le \exp\left( -\frac{\delta^2 nq}{2+\delta} \right).
    \]
\end{lemma}

\end{document}